\documentclass[submission,copyright,creativecommons]{eptcs}
\providecommand{\event}{AiSoS 2013} 
\usepackage{breakurl}             
\usepackage{xcolor,colortbl}
\usepackage[normalem]{ulem}

\title{System-of-Systems Complexity}
\author{Hermann Kopetz
\institute{Institut f\"{u}r Technische Informatik\\
Vienna University of Technology\\
Vienna, Austria}
\email{H.Kopetz@gmail.com}
}
\def\titlerunning{System-of-Systems Complexity}
\def\authorrunning{Hermann Kopetz}
\begin{document}
\maketitle

\begin{abstract}
The global availability of communication services makes it possible to
interconnect independently developed systems, called constituent systems, to provide
new synergistic services and more efficient economic processes. The characteristics of
these new Systems-of-Systems are qualitatively different from the classic monolithic
systems. In the first part of this presentation we elaborate on these differences,
particularly with respect to the autonomy of the constituent systems, to dependability,
continuous evolution, and emergence. In the second part we look at a SoS from the
point of view of cognitive complexity. Cognitive complexity is seen as a relation
between a model of an SoS and the observer. In order to understand the behavior of a
large SoS we have to generate models of adequate simplicity, i.e, of a cognitive
complexity that can be handled by the limited capabilities of the human mind. We will
discuss the importance of properly specifying and placing the relied-upon message
interfaces between the constituent systems that form an open SoS and discuss
simplification strategies that help to reduce the cognitive complexity.
\end{abstract}

\section{Characteristics of a System-of-Systems}
In an increasing networked society many new super-systems are developed by the interconnection of existing legacy systems. This new type of systems, called {\it System-of-Systems}, is fundamentally different from the classic monolithic system, as depicted in Table~\ref{table1}.

\begin{table}[h]
\begin{center}
\begin{tabular}{rcc}
{\bf Characteristic}      & {\bf Monolithic}     & {\bf SoS}        \\
Scope of System           & Fixed (known)        & Not known        \\
Structure                 & Hierarchical         & Networked        \\
Requirements and Spec.    & Fixed                & Changing         \\
Control                   & Central              & Autonomous       \\
Evolution                 & Version control      & Uncoordinated    \\
Testing                   & Test phases          & Continuous       \\
Implementation technology & Given and fixed      & Unknown          \\
Faults (Physical, Design) & Exceptional          & Normal           \\
Emergence                 & Insignificant        & Important        \\
System development        & Process model        & ???              \\
\end{tabular}

\caption{Monolithic Systems versus System of Systems (adapted from \cite{Maier1998})\label{table1}}
\end{center}
\end{table}

Whereas a monolithic system is in the \textit{sphere of control}, i.e., the \textit{governance}, of a single organization,
the constituent systems (CSs) of an SoS belong to different organization with different organizational
objectives. From the point of view of a SoS, its CSs are thus autonomous and cannot be \textit{forced to}
contribute to the overall goal of the SoS, they can only be \textit{influenced to} contribute by providing proper
incentives and reward structures. Since the CSs belong to different organization adhering to different
architectural styles, the information that is exchanged across an interface will normally be based on
different syntax and semantics, thus leading to \textit{property mismatches} at the interfaces. These property
mismatches, both at the syntactic and semantic level, must be resolved in order that a meaningful
communication among the different CSs can be established.

Whereas the internal structure of a CS can be mapped into a hierarchy, the structure of an SoS is more
likely to be a mesh, implying that a hierarchical decomposition of an SoS is in general not possible. In
order to control the cognitive complexity of SoS models, an \textit{aspect-oriented approach} is followed.
Only those aspects of the CSs that are relevant for the purpose of the integration into the SoS are visible at the relied upon messages interfaces (RUMI) between CSs, thus reducing the amount of
information that needs to be dealt with at an interface in order to understand the behavior of the SoS.
The deliberate placement of the RUMIs and the precise specification of their syntax, semantics and
temporal properties are of utmost importance in the design of an SoS.

Every successful system that is embedded in the real world must continuously evolve in order to
remain relevant for its users. An \textit{open system} that does not adapt to the ever-changing requirements of
our highly dynamic world will soon become obsolescent. From the point of evolution, monolithic
systems and SoSs are fundamentally different. Whereas the version control in a monolithic system
ensures that all changes are consistent before a new version of a monolithic system is released, the
evolutions of the different CSs forming an SoS generally cannot not be coordinated in this way. A CS
is changed whenever there is a \textit{need to change} required by the owning organization, hardly
considering or coordinating all the possible consequences of the changes on the overall SoS behavior.
This puts many of the well-established system engineering principles and design methods up for
discussion. A static authoritative specification of an SoS does not exist. The same system that is
correct today may not be correct tomorrow, since the world has changed.

The interactions of the CSs of an SoS can lead to the appearance of unique properties at the SoS level
that cannot be attributed to any of the properties of the CSs. These new properties are called \textit{emergent
properties}. Emergent properties are novel, irreducible, and holistic---they disappear when the system
is partitioned into its subsystem. Consider the example of \textit{deadlock} in a distributed computer systems.
Emergent properties can be \textit{unforeseen} or \textit{expected}, they can be \textit{beneficial} or \textit{detrimental}. At its first
appearance, emergent properties are often unforeseen. At the moment, the general issues revolving
around the concept of emergence are not well understood and it is a challenge to detect and avoid
detrimental emergence properties in a new SoS.

The objective of SoS design is the establishment of a framework that supports unforeseen changes---
this is major paradigm shift in our industry. Understanding the proper handling of the evolution of an
SoS is thus a most relevant theme for practitioners and researchers.

\section{Cognitive Complexity}

According to the Merriam Webster Dictionary \cite{mw2013} \textit{complex} means \textit{hard to separate, analyze or solve;
having many parts or aspects that are usually interrelated}. We can classify complexity as follows:
\begin{itemize}
  \item Complexity as a Property of a scenario
    \begin{itemize}
      \item[$\circ$] \textit{Structural Complexity} that is concerned with the topology of the parts and the links
among the parts.
      \item[$\circ$] \textit{Dynamic Complexity} that is concerned with the behavior of the parts and their
dynamic interactions, such as causality, feedback or delayed response.
    \end{itemize}
  \item Complexity as a Relation
    \begin{itemize}
     \item[$\circ$] \textit{Cognitive Complexity:} Relation between a scenario and an observer.
     \item[$\circ$] \textit{Socio Political Complexity:} Relation between a scenario and society.
  \end{itemize}
\end{itemize}

In this presentation we focus on Cognitive Complexity (the antonym of simplicity) of a SoS, which is a
relation between a scenario and an observer who tries to understand the scenario. We understand the
world around us by conceptual modeling, i.e., by the generation of a hierarchy of models of reality that
are agreeing with the cognitive capabilities of the human mind. Understanding means that the
concepts and dependencies used to represent a model are adequately linked with concepts already
familiar to the observer. The closer these links, the better the understanding. We consider the elapsed
time needed to understand a model by an observer of the intended group of observers as a feasible
measure for the cognitive complexity of a model.
A conceptual model is an abstraction that is formed for the purpose of understanding a chosen aspect
of the scenario, such as: structure, behavior, timeliness, dependability, etc.. If the purpose of a model
is not crystal clear, it is not possible to construct a simple model of a scenario because it cannot be
decided what is relevant and what is irrelevant (and can be neglected) when constructing an abstract
model for the specified purpose. Take the example of celestial mechanics: If the purpose of the model
is the understanding of the movement of the heavenly bodies, we abstract from the whole diversity of
the world and reduce it to a mass point.
In the context of a SoS, understanding the behavior is of utmost importance. The complexity of a
model of behavior of a SoS depends on the static and dynamic properties of the constituent systems,
the organization of the SoS (i.e., the static structure and dynamic interaction of the CSs) and the
experience of the observer in dealing with such an SoS. In order to ease the understanding it may be
necessary to construct a hierarchy of behavioral models, where at the lower level a model is a
refinement of a higher-level model. Each model must take account of the limits of human cognition—
at most five plus minus two chunks of information can be represented in short term memory \cite{miller1956} and
humans are not capable to handle relations with more than four variables \cite{halford2005many}.

\section{Relied-Upon Message Interfaces (RUMI)}
As a rule the CSs of a SoS interact by the exchange of messages only. The internal architecture of an
SoS is determined by the placement and specification of the Relied Upon Message Interfaces (RUMIs)
among the CSs. A RUMI should be a stable interface that establishes the boundaries between two
interacting CSs by specifying the messages that are exchanged between these CSs. RUMIs must be
fully specified w.r.t. their syntax, semantics and temporal behavior. Whereas the syntactic
specification deals with the structure of the interface and establishes the form of the syntactic units, the
data items at the interface, the semantics specification assigns meaning to these interface data items.
The semantic specification consists of an interface model that explains the data items by using
concepts that are familiar to the user of the interface. Since the two CSs that meet at a RUMI are
normally designed by different organizations that use different architectural styles, there will be two
different semantic specifications of the same RUMI, depending from which side the RUMI is viewed.
The temporal specification of a RUMI must outline the temporal properties of the message exchanges.

\section{Simplification Strategies}
The major challenge of information system design is the building of a software/hardware/people
artifact that provides the intended service under given constraints and where relevant properties of this
artifact (e.g., the behavior) can be modeled at different levels of abstraction by models of adequate
simplicity. The following design principles, developed for the control of the cognitive complexity of
monolithic systems, are also relevant for Systems-of-Systems \cite[p.~46]{kopetz2011real}:
\begin{itemize}
\item {\it Principle of Abstraction:} The behavior of a large system can be explained by a hierarchy of
models, where each model considers the limited cognitive capability of the human mind, as
explained in Section 3.

\item {\it Principle Separation of Concern:} This principle helps to build simple systems by
disentangling functions that are separable in order that they can be grouped in self-contained
architectural units,

\item {Principle of Causality:} The analytical-rational problem solving subsystem of humans excels
in reasoning along causal chains. The deterministic behavior of basic mechanisms makes it
possible that a causal chain between a cause and the consequent effect can be established
without a doubt. Probabilistic dependencies between cause and effect are more difficult to
grasp.

\item {\it Principle of Segmentation:} This principle suggests that hard-to-understand behavior should
be decomposed, wherever possible, into a serial behavioral structure such that a sequential
step-by-step analysis of the behavior becomes possible.

\item {\it Principle of Observability:} Non-visible communication channels among architectural units
pose a severe impediment for the understanding of system behavior. This can be avoided by
supporting a multicast topology in the basic message passing primitive. It is then possible to
observe the external behavior of any component without a probe effect.

\item {\it Principle of a Consistent Global Time:} This principle suggests that a sparse global time base
should be introduced in all CSs of an SoS such that system-wide consistent temporal relations
(e.g., simultaneity) and physical temporal distances among events can be established on the
basis of global time-stamps.

\end{itemize}

In addition, the following specific design principles should help to reduce the cognitive complexity of
System of Systems:

\begin{itemize}

\item {\it Principle of Classification of Expected Changes:} In the context of evolution of an SoS we
distinguish between minor and major changes: minor changes are confined to the internals of
a CS and have no effect on a RUMI. Major changes have an effect on one or more RUMIs. In
a large SoS is advantageous to categorize RUMIs (and consequently changes) w.r.t. their
impact on the overall SoS architecture on an even finer scale.

\item {\it Principle of Outside Flexible, Inside Stable Interfaces:} The interfaces between a cyber-system
and its external environment are subject to the evolution of the external world. This evolution
is out of control of the cyber-system. Internal relied upon message interfaces (RUMI) can
only be controlled, if both sides of the interface are in the sphere of control of the system
designer. It is therefore good practice to provide a gateway component between an internal
RUMI and an interface to the external world.

\item {\it Principle of Intrinsic vs. Extrinsic Complexity:} Extrinsic complexity is concerned with the
service of a CS at a given relied upon message interface (RUMI). Intrinsic complexity is
concerned with the design of the internals a CS. From the SoS point of view, a low extrinsic
complexity of the RUMIs of the CSs should be strived for. In many cases, a low extrinsic
complexity is achieved at the price of a high intrinsic complexity.

\item {\it Principle of Specifying Goals, not Processes:} It is much simpler to specify a goal state—a
 problem solution— than to specify a process that leads from the current state to the goal state
  \cite{Newell1972}. In many SoSs, a top layer—the federation layer—interfaces with the problem owner and
     partitions the user’s goal into sub-goals. Selected CSs are activated to find solutions to the
      sub-goals specified in the respective RUMIs. A CS should be autonomous in finding a
     solution to a sub-goal that is specified in its RUMI. This process can be recursive.

\item {\it Principle of Autonomic Fault Mitigation:} Considerations about fault containment and the
control of error propagation have a decisive influence on the placement of the RUMIs. A CS
should form an encapsulated fault-containment unit (FCU). Error detection mechanisms must
be provided in the SoS to detect failures of a CS within a short error detection latency. A CS
should be capable to recover from a transient fault within a defined error recovery time. This
requires the provision of appropriate recovery points as part of the design.

\item Principle of Isomorphic Decomposition: An SoS can modeled from different viewpoints,
such as behavior, fault containment, evolution, maintenance, etc. Each viewpoint can be
explained by a hierarchy of models. Ideally, the analysis of a SoS according to these different
viewpoints should result in the same decomposition, which is then called an isomorphic
decomposition \cite{wimsatt1975}. The design of an isomorphic structure is an art that requires experience
and foresight.

\end{itemize}

\section{Conclusion}
Systems of systems are substantially different from monolithic systems---many of the established
design methods need to be revisited. Since the substantial cognitive effort required to understand a
system-of-system from the different perspectives is the main cause for the massive engineering effort
in design and operation, it is a worthwhile goal to structure a System of System such that the cognitive
complexity for understanding the designed artifact is reduced.

\nocite{*}
\bibliographystyle{eptcs}
\bibliography{generic}
\end{document}